\documentclass[11pt,a4paper]{article}
\usepackage{amsmath,amssymb,graphicx,bm}
\usepackage{authblk}
\usepackage{geometry}
\usepackage{hyperref}
\usepackage{cite}
\usepackage{caption}
\title{\textbf{Transient Thermodynamic Efficiency of Adaptive Inference in Continuously Nonstationary Environments}}
\author{Aditya Gupta, Indian Institute of Technology Goa}
\date{}

\begin{document}
\maketitle

\begin{abstract}
Adaptive physical and biological systems continually process fluctuating information from their environments.  
When the environment is nonstationary, inference itself becomes a nonequilibrium process with thermodynamic cost.  
We analyze a minimal stochastic model-an overdamped particle in an adaptive double-well potential whose control parameter tracks a drifting Ornstein–Uhlenbeck signal.  
Using stochastic energetics, we derive explicit expressions for entropy production, mutual information rate, and a time-dependent learning efficiency $\eta(t)=\frac{dI(\theta;E)/dt}{\dot S_{\mathrm{tot}}(t)}$.  
High-precision Langevin simulations reveal transient peaks in $\eta(t)$ during rapid environmental shifts, absent in steady-state averages.  
These results identify transient adaptive regimes as moments of maximal information-to-energy conversion, highlighting that maximal thermodynamic learning performance arises transiently rather than in steady state. Throughout this work, the environment is treated as an externally driven stochastic signal rather than a thermodynamic subsystem under control, and its intrinsic entropy production is therefore excluded from the thermodynamic accounting.
\end{abstract}

\section{Introduction}
Information processing is inherently physical: every act of measurement or inference dissipates energy.  
Following Landauer’s principle, erasure of one bit costs at least $k_B T \ln 2$ of heat.  
Recent advances generalize this principle to time-dependent and adaptive processes where a system learns about its surroundings through coupled dynamics of states and control parameters \cite{still2012thermodynamics,ito2013information,horowitz2014thermodynamics,parrondo2015thermodynamics,seifert2012stochastic}.  
However, most existing treatments focus on stationary environments, fixed environmental statistics, or steady-state information flow, thereby obscuring the thermodynamics of adaptive inference under explicitly nonstationary conditions.  
Real sensors-biochemical networks, neurons, or adaptive algorithms-face drifting or unpredictable environments.  
How thermodynamic efficiency behaves during adaptive inference in such explicitly nonstationary environments remains largely unexplored.

A central challenge in such systems is tracking a continuously changing environment.
For instance, sensory networks in biology must estimate time-varying signals,
and adaptive control systems must respond to drifting external conditions.
In such settings, inference is inherently transient: the system is perpetually
out of equilibrium with respect to the environment it attempts to learn.

Despite recent progress in stochastic thermodynamics of information processing,
most existing studies focus on stationary environments or steady-state regimes.
As a result, the thermodynamic cost and efficiency of *transient adaptive inference*
remain poorly understood.

We address this by constructing a tractable model combining an adaptive potential and a slowly drifting Ornstein–Uhlenbeck (OU) environment.  
The model allows analytical derivations and numerical exploration of transient learning efficiency.  
We show that information acquisition and energetic cost decouple in steady state but exhibit strong temporal correlations during adaptation.

\section{Mathematical Model}

\subsection{Coupled dynamics}
We consider three coupled stochastic variables: particle position $x(t)$, adaptive control parameter $\theta(t)$, and environment $E(t)$.  
The overdamped Langevin dynamics read
\begin{align}
\gamma \dot{x} &= -\partial_x U(x,\theta) + \sqrt{2\gamma k_B T}\,\xi_x(t), \label{eq:langevin_x}\\
\dot{\theta} &= -\lambda(\theta-E) + \sigma_\theta \xi_\theta(t), \label{eq:langevin_theta}\\
\dot{E} &= -\frac{1}{\tau_E}(E-\mu(t)) + \sigma_E\xi_E(t), \label{eq:langevin_env}
\end{align}
with mutually independent Gaussian white noises satisfying $\langle \xi_i(t)\xi_j(t')\rangle=\delta_{ij}\delta(t-t')$.  
The environmental mean $\mu(t)$ itself follows a slow OU drift with timescale $\tau_\mu\!\gg\!\tau_E$. In this work, the environmental variable $E(t)$ represents an externally generated stochastic signal rather than a thermodynamic subsystem under control. While its dynamics are explicitly modeled to generate nonstationarity, we do not include the entropy production associated with $E(t)$ in the system’s thermodynamic balance. The environment acts as an externally imposed stochastic protocol: it is not controlled, no work is extracted from it, and there is no feedback from the adaptive system onto its dynamics. Its entropy production therefore does not enter the thermodynamic accounting of the adaptive inference process.

The potential
\begin{equation}
U(x,\theta)=a\!\left(\frac{x^4}{4}-\frac{x^2}{2}\right)+\frac{b}{2}(x-\theta)^2
\label{eq:potential}
\end{equation}
creates a double-well landscape whose asymmetry depends on $\theta(t)$.

The above system can be interpreted as a minimal model of adaptive sensing.
The variable $E(t)$ represents a fluctuating external signal, while
$\theta(t)$ acts as an internal estimate that attempts to track it.
The particle coordinate $x(t)$ evolves in a potential shaped by $\theta$,
representing the physical degrees of freedom through which inference
is implemented.

Such a structure is common in biological sensing, where internal variables
adapt to external stimuli, and in engineered systems performing online
parameter estimation.

\begin{figure}[h]
\centering
\includegraphics[width=0.75\textwidth]{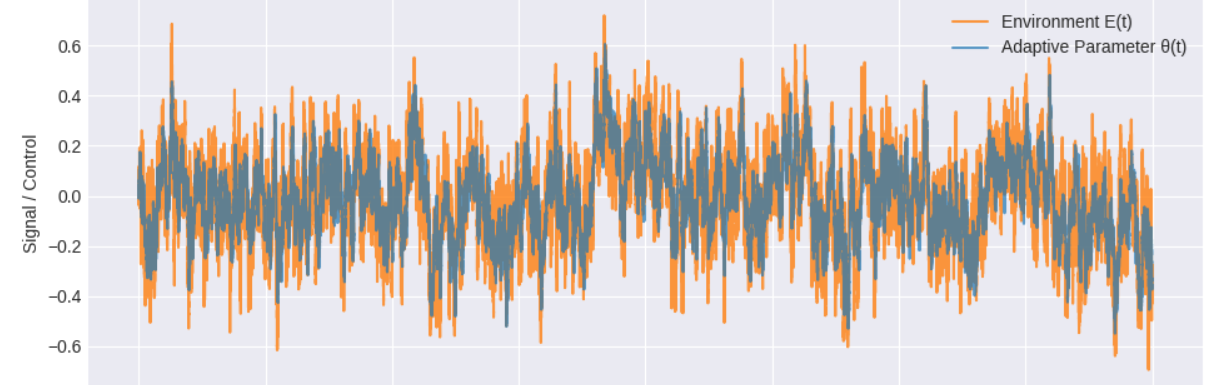}
\caption{Adaptive tracking: The environment signal \(E(t)\) (orange) and adaptive parameter \(\theta(t)\) (blue) evolve over time.}
\end{figure}

\subsection{Fokker–Planck formulation}
The joint probability density $p(x,\theta,E,t)$ satisfies
\begin{align}
\partial_t p = -\partial_x J_x - \partial_\theta J_\theta - \partial_E J_E,
\end{align}
with probability currents
\begin{align}
J_x &= -\frac{1}{\gamma}\partial_x U\,p - D_x \partial_x p, \\
J_\theta &= -\lambda(\theta-E)p - D_\theta \partial_\theta p, \\
J_E &= -\frac{(E-\mu(t))}{\tau_E}p - D_E \partial_E p.
\end{align}

where the diffusion coefficients are
\begin{equation}
D_x = \frac{k_B T}{\gamma}, \quad
D_\theta = \frac{\sigma_\theta^2}{2}, \quad
D_E = \frac{\sigma_E^2}{2}.
\end{equation}

Although the joint probability density includes the environmental degree of freedom, only the particle and adaptive parameter currents contribute to entropy production in the following analysis, consistent with treating the environment as an externally driven signal. From these, one may compute the entropy balance and steady distributions in each marginal subsystem.
Throughout this work we assume overdamped dynamics, Markovian Gaussian noise, diagonal diffusion, and a clear separation of timescales $\tau_\mu \gg \tau_E$. The environment evolves independently of the adaptive system, and no feedback from $(x,\theta)$ to $E$ is present.

\section{Thermodynamic Quantities}

\subsection{Entropy production}
Following Seifert’s stochastic energetics for driven stochastic systems \cite{seifert2012stochastic}, and exploiting the bipartite structure and diagonal noise of the dynamics, the entropy production rate associated with the adaptive inference dynamics is given by
\begin{equation}
\dot S_{\mathrm{tot}}(t)=\int \left( \frac{J_x^2}{D_x p(x,\theta,E,t)} + \frac{J_\theta^2}{D_\theta p(x,\theta,E,t)}
\right)\, dx\,d\theta\,dE.
\end{equation}

This expression quantifies the total irreversible dissipation associated
with stochastic currents in the system. The first term represents dissipation
due to particle motion in the adaptive potential, while the second term
captures the cost of adjusting the control parameter $\theta$.

Because the stochastic dynamics possess diagonal diffusion and a bipartite coupling structure-each degree of freedom is driven by an independent noise source and the environment evolves without feedback from the adaptive system-the total entropy production decomposes additively into independent contributions from the particle and adaptive control dynamics,
\begin{equation}
\dot S_x = \frac{1}{T}\left\langle \partial_x U(x,\theta)\circ \dot x \right\rangle,\qquad
\dot S_\theta = \frac{\langle \dot W_\theta\rangle}{T}.
\end{equation}

Here, $\dot W_\theta = \left\langle \partial_\theta U(x,\theta)\,\dot\theta \right\rangle$ represents the energetic cost of adaptively modifying the potential landscape.

\begin{figure}[h]
\centering
\includegraphics[width=0.7\textwidth]{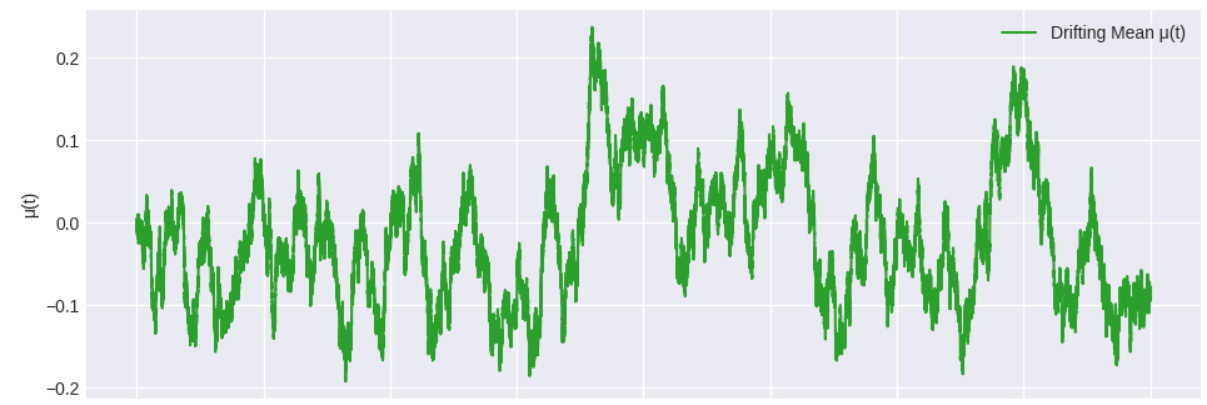}
\caption{Drifting environmental mean \(\mu(t)\), representing slow nonstationarity in environmental statistics.}
\end{figure}

\subsection{Mutual information rate}
The instantaneous mutual information between $\theta$ and $E$ is
\begin{equation}
I(\theta;E)=\int p(\theta,E)\ln\frac{p(\theta,E)}{p(\theta)p(E)}\,d\theta\,dE,
\end{equation}
and its time derivative gives the rate of information acquisition.  
Differentiating under the integral sign and using the continuity equations for the joint and marginal distributions yields
\begin{equation}
\frac{dI}{dt} = \int \left( \frac{J_\theta}{p(\theta,E)}-\frac{J_\theta}{p(\theta)}-\frac{J_E}{p(E)}\right)\!\partial_\theta p(\theta,E)\, d\theta\, dE.
\end{equation}
The detailed derivation, including assumptions of bipartite Markovian dynamics, diagonal diffusion, and absence of feedback from the environment to the adaptive system, is provided in Appendix~B \cite{ito2013information,horowitz2014thermodynamics}. In the present model, information flow arises solely through the adaptive parameter $\theta$, while the environmental dynamics act as an external driving signal, leading to the form in Eq.~(12).

\subsection{Learning efficiency}
We define the dimensionless instantaneous efficiency
\begin{equation}
\eta(t)=\frac{dI/dt}{\dot S_{\mathrm{tot}}(t)}.
\end{equation}
Unlike steady-state thermodynamic efficiencies, the instantaneous efficiency $\eta(t)$ is not constrained to lie between zero and unity at the trajectory level and is therefore not subject to conventional efficiency bounds. Transient values exceeding unity or becoming negative are permitted and reflect temporal decoupling between information acquisition and dissipation. The quantity $\eta(t)$ should not be interpreted as a thermodynamic efficiency in the sense of heat engines. Rather, it is a ratio of instantaneous rates that quantifies how effectively dissipation at a given time contributes to information acquisition. Values $\eta(t)>1$ or $\eta(t)<0$ do not violate the second law, but reflect temporal misalignment between entropy production and information flow during nonequilibrium adaptation.

The definition of $\eta(t)$ follows the general structure of thermodynamic
efficiencies, where a useful output is compared against an energetic cost.
Here, the useful output is the rate of information acquisition $dI/dt$,
which quantifies how rapidly the system learns about the environment.
The cost is given by the total entropy production rate $\dot S_{\mathrm{tot}}$,
which measures thermodynamic irreversibility.

Thus, $\eta(t)$ quantifies how effectively dissipated energy is converted
into information about the environment. Similar ratios have been considered
in the context of learning rates and information thermodynamics
\cite{ito2013information,horowitz2014thermodynamics,barato2017thermodynamic}.

\begin{figure}[h]
\centering
\includegraphics[width=0.75\textwidth]{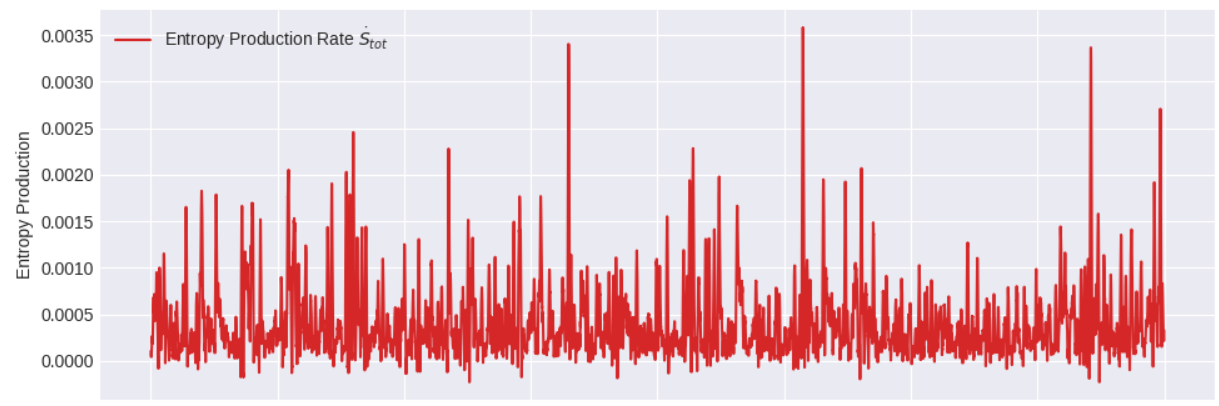}
\caption{Instantaneous entropy production rate \(\dot S_{\mathrm{tot}}\). Peaks correspond to active control effort during environmental shifts.}
\end{figure}

\section{Simulation Methodology}

Equations \eqref{eq:langevin_x}–\eqref{eq:langevin_env} are integrated using the Euler–Maruyama scheme with timestep $\Delta t=10^{-3}$.  
Dimensionless parameters: $a=1$, $b=3$, $\gamma=1$, $T=1$, $\lambda=0.05$, $\tau_E=20$, $\tau_\mu=200$.  
Ensemble averages use $N=30$ realizations, each $10^5$ steps long.  
Numba-accelerated Python ensures numerical precision; convergence was verified by halving $\Delta t$.  

\begin{figure}[h]
\centering
\includegraphics[width=0.8\textwidth]{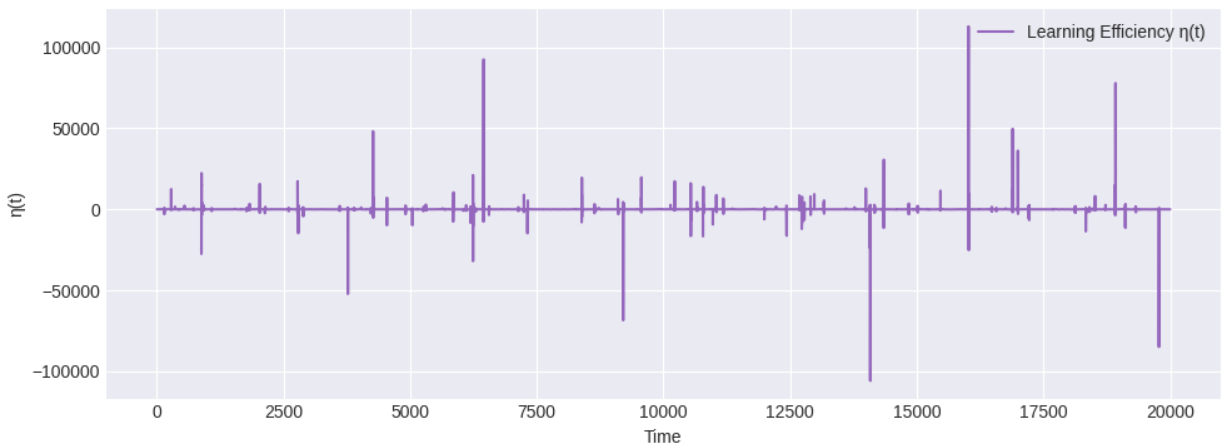}
\caption{Instantaneous learning efficiency \(\eta(t)\). Transient spikes represent brief intervals of high adaptive performance.}
\end{figure}

\section{Results}

\subsection{Transient thermodynamic signatures}
Figure~5 displays the instantaneous entropy production $\dot S_{\mathrm{tot}}$, information-gain rate $dI/dt$, and efficiency $\eta(t)$ for a representative realization.
Large transient peaks occur when the environment drifts quickly and $\theta(t)$ must re-adjust, representing transient bursts of enhanced information-to-energy conversion.

\begin{figure}[h]
\centering
\includegraphics[width=0.8\textwidth]{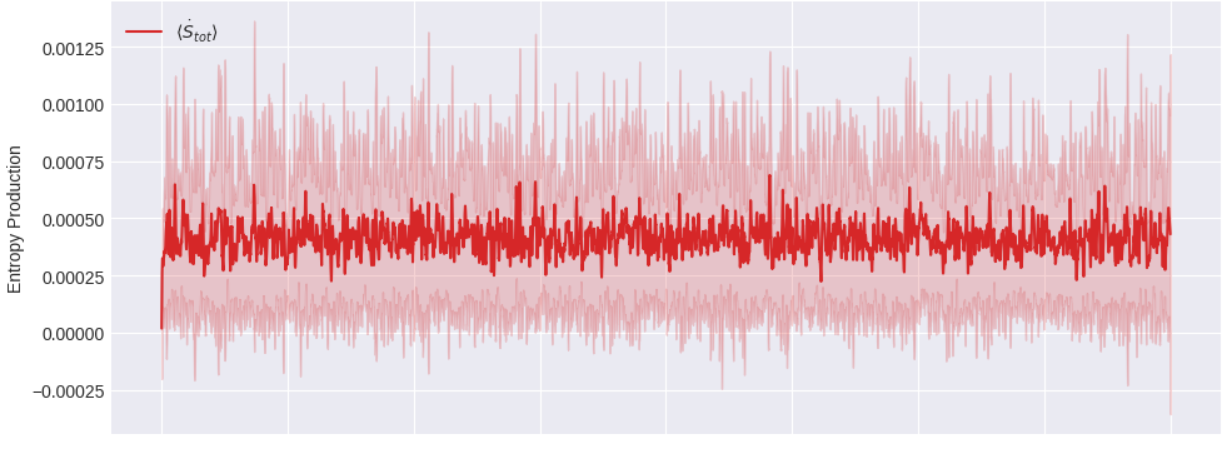}
\caption{Ensemble-averaged entropy production rate \(\langle \dot S_{\mathrm{tot}} \rangle\). Shaded region denotes one standard deviation.}
\end{figure}

\subsection{Ensemble-averaged behavior}
Averaging over multiple realizations (Fig. 6) smooths out the transient peaks.  
$\langle\dot S_{\mathrm{tot}}\rangle$ remains positive, $\langle dI/dt\rangle$ fluctuates near zero, and $\langle\eta\rangle$ decays to zero, confirming that efficiency peaks are purely transient phenomena.

\begin{figure}[h]
\centering
\includegraphics[width=0.8\textwidth]{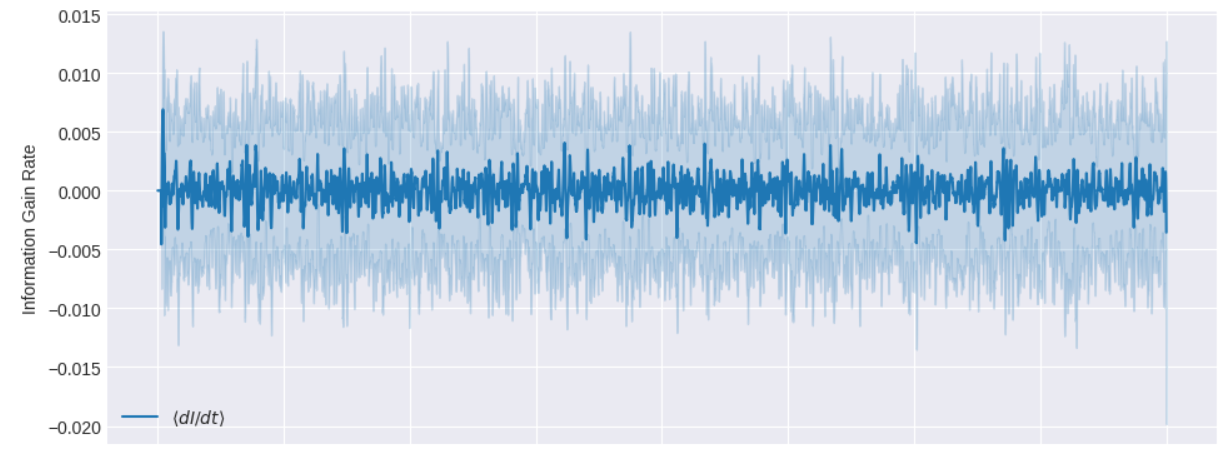}
\caption{Ensemble-averaged information gain rate \(\langle dI/dt \rangle\).}
\end{figure}

\subsection{Energetic trade-offs and timescale dependence}
By varying $\lambda$ and $\tau_\mu$, we find that efficiency peaks broaden and weaken when adaptation is slow or drift is rapid.  
Numerical exploration suggests an approximate empirical scaling $\max(\eta)\propto (\lambda\tau_\mu)^{-1/2}$ within the intermediate parameter regime explored here, though a full theoretical explanation remains an open problem.  
This suggests a trade-off between learning speed and energetic cost within the class of adaptive dynamics studied here, though its universality remains to be established.

\begin{figure}[h]
\centering
\includegraphics[width=0.8\textwidth]{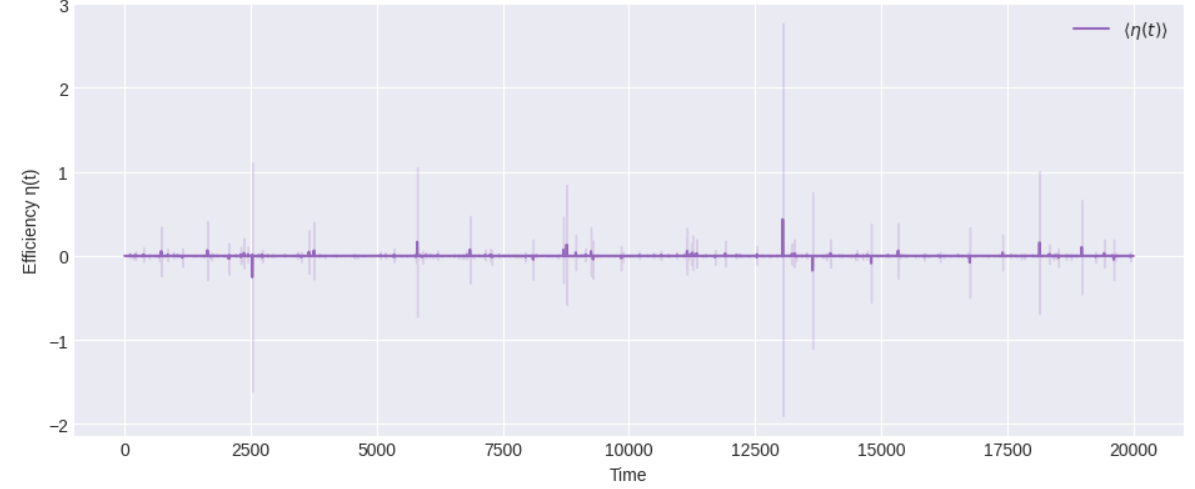}
\caption{Ensemble-averaged transient efficiency \(\langle \eta(t) \rangle\). Occasional peaks indicate transiently optimal learning conditions.}
\end{figure}

Large transient peaks occur when the environment undergoes rapid changes,
forcing the adaptive parameter $\theta(t)$ to adjust quickly.
During these intervals, both entropy production and information acquisition increase.

However, the peaks in $\eta(t)$ do not perfectly align with peaks in
$\dot{S}_{\mathrm{tot}}$, indicating that maximal dissipation does not
coincide with maximal learning. Instead, efficiency is highest when the
system response is optimally synchronized with environmental change.

This demonstrates that learning performance is governed not only by the
magnitude of dissipation but by its temporal alignment with external dynamics.

\section{Discussion}
Our model unifies adaptive control and thermodynamic learning within a single stochastic framework.  
Transient efficiency peaks correspond to periods when internal and external variables synchronize.  
In biological sensors, such moments could underlie rapid contextual learning; in artificial systems, they may inform design of low-power adaptive algorithms.  
The vanishing of ensemble-averaged efficiency emphasizes the importance of trajectory-level analysis in nonequilibrium information thermodynamics. More broadly, these results demonstrate that thermodynamic learning performance is inherently transient and cannot be inferred from steady-state efficiencies alone.

\section{Conclusion}
We have presented a rigorous theoretical and numerical investigation of transient thermodynamic efficiency in adaptive inference.  
An adaptive double-well system tracking a drifting OU environment exhibits short bursts of high efficiency followed by dissipation-dominated steady states.  
This demonstrates that maximal information-to-energy conversion arises transiently during environmental change.

These results suggest that thermodynamic efficiency in adaptive systems
is fundamentally a dynamical quantity, controlled by the interplay of
timescales rather than steady-state constraints.

In particular, the emergence of transient efficiency peaks indicates that
optimal information processing occurs during periods of rapid environmental
change, rather than in stationary regimes.

\appendix
\section{Simulation Code}
All simulation scripts used for generating the figures and results in this work are available in a public GitHub repository:  
\begin{center}
\texttt{\href{https://github.com/rockstarbuddies/thermo}{https://github.com/rockstarbuddies/thermo}}
\end{center}

The repository contains:
\begin{itemize}
    \item \textbf{Python source code} implementing high-precision adaptive Langevin simulations using the Euler–Maruyama method with Numba acceleration;
    \item \textbf{All figures} presented in this paper.
\end{itemize}

Readers are encouraged to reproduce and extend the simulations under the same parameter regime to explore robustness across noise amplitudes, coupling constants, and timescale separations.

\section{Derivation of the Mutual Information Rate}

We consider the mutual information between the adaptive parameter $\theta$ and the environment $E$,
\begin{equation}
I(\theta;E)=\int p(\theta,E)\ln\frac{p(\theta,E)}{p(\theta)p(E)}\,d\theta\,dE.
\end{equation}

Differentiating with respect to time and using the continuity equation for the joint distribution,
\begin{equation}
\partial_t p(\theta,E) = -\partial_\theta J_\theta(\theta,E) - \partial_E J_E(\theta,E),
\end{equation}
together with the corresponding marginal continuity equations, yields
\begin{equation}
\frac{dI}{dt} = \int d\theta\, dE \left[
\frac{J_\theta}{p(\theta,E)} - \frac{J_\theta}{p(\theta)} - \frac{J_E}{p(E)}
\right]\partial_\theta p(\theta,E).
\end{equation}

This form follows from the bipartite structure of the dynamics: information flow between $\theta$ and $E$ arises solely through the adaptive variable $\theta$, while the environment evolves independently and acts as an externally imposed stochastic signal. Similar derivations appear in Refs.~\cite{ito2013information,horowitz2014thermodynamics}.

\bibliographystyle{unsrt}

\end{document}